# Chapter 5

# Predictive analytics using Social Big Data and machine learning

Bilal Abu-Salih[1], Pornpit Wongthongtham[2]
Dengya Zhu[3] , Kit Yan Chan[3] , Amit Rudra[3]

[1]The University of Jordan
[2] The University of Western Australia
[3] Curtin University

**Abstract:** The ever-increase in the quality and quantity of data generated from day-to-day businesses operations in conjunction with the continuously imported related social data have made the traditional statistical approaches inadequate to tackle such data floods. This has dictated researchers to design and develop advance and sophisticated analytics that can be incorporated to gain valuable insights that benefit business domain. This chapter sheds the light on core aspects that lay the foundations for social big data analytics. In particular, the significant of predictive analytics in the context of SBD is discussed fortified with presenting a framework for SBD predictive analytics. Then, various predictive analytical algorithms are introduced with their usage in several important application and top-tier tools and APIs. A case study on using predictive analytics to social data is provided supported with experiments to substantiate significance and utility of predictive analytics.

## 5.1 Introduction

The abundance of SBD gives organisations the opportunity to maximise their use of a wealth of available information to increase their revenues. Hence, there is an urgent need to capture, load, store, process, analyse, transform, interpret, and visualise a diversity of social datasets to develop meaningful insights that are specific to an application's domain. In this context, companies incorporate advanced social data analytics when designing effective marketing strategies and seek to leverage the interactive quality of online social services. Thus, to create the required interaction with their customers, companies use many modern communication to attract customers and visitors to their online social platforms. Consequently, it is necessary for companies to analyse their customers' social content and classify the customers into appropriate categories based on for example their topics of interest, to deliver the right message to the right category. This ultimate goal should harness technical solutions with a capacity to infer the meaning of social content at the user level and post level [1-5].

In the previous chapter an ontology-based approach is developed to semantically analyse the social data at two levels, i.e. the entity level and the domain level, thus extract semantics of textual data and define the domain of data. This approach has shown successful in supporting and boosting the resultant output of semantics analytic providers (i.e. IBM Watson NLU). However, there is a need to extend this work to provide a platform for automatically classify and predict the domain of interest at the user level. In particular, the existing approaches to topic extraction, modelling and classification rely on statistical bag-of-words techniques such as Latent Dirichlet Allocation(LDA) [6]. These techniques have shown several limitations include: (i) the number of designated topics is set fixed and should be known before the analysis [7]; (ii) the topics mined by these models do not contemplate the temporal aspects [8]; (iii) these models are considered as monolingual topic models, hence they do not differentiate idioms of the same language [9] and; (iv) these models are unable to infer high-level topics particularly from short text such as tweets [10].

This chapter is envisioned to address this research issue and to extend the endeavour of the previous chapter. This is by means of leveraging the external semantic web knowledge bases and machine learning modules to mitigate the disambiguation in the textual content and classify and predict the domain of interest at the user as it will be illustrated in the next sections.

The structure of this chapter starts with outlining the key components of the commonly used modelling framework for SBD. Then, a brief introduction is given to various machine learning algorithms for predictive analytics. In the following section, we discuss a selective set of applications that benefit from the predictive analytics as well as presenting a selected set of well-known tools and APIs. A case study is then provided on social politics domain incorporating various machine learning and classification algorithms.

## 5.2 Predictive modelling framework for Social Big Data

Constructing an efficient framework to analyse and infer value form SBD is not an easy exercise. In fact, it is a cohesive process requires a careful consideration into each step in this process to insure the outcomes of each step benefit the subsequent ones. In spite of the fact that the detailed analysis might differ based on the nature of each addressed problem, there is a generic paradigm that is commonly followed in conducting predictive modelling, particularly those pertaining to SBD. One of the crucial steps in this paradigm is to define and formulate a problem adequately before perusing the analysis. As psychologist and philosopher John Dewey said that “a problem well-stated is half-solved” [11]. This indeed economizes tremendous amount of effort and facilitate for the analyst carrying out the following steps in a correct fashion. Understanding a problem embodies indicating the objective, anticipated outcomes and the methodology to resolve it. It also includes a well understanding to the business requirements so analyst can frame the right approach to solve it [12-14].

SBD provides an important arena for companies to obtain an added value that benefits business domain. This is though market analysis, listening to the Voice of Customer (VoC) and for sentiment analysis to feed Business Intelligence applications [15]. Therefore, a methodology is required to infer the trustworthiness of unstructured data from different SBD sources such as social media networks, news agencies, and web logs, and to store a collection of trustworthy unstructured data using the existing data warehousing solutions. In a related context, semantically enriching textual data and imposing structure on the unstructured nature of SBD when transferring these to data warehouses is another significant task. Discovering the semantics of social data will enhance the quality and accuracy of data stored in data warehouses which will dramatically affect the decision-making process as well as the quality of extracted reports. The application of semantic web technology resolves the issue of the ambiguity of data and provides metadata which helps related data to be understood and interpreted accurately. Meanwhile, ontology is utilised to define and collect semantically-related concepts and relations between concepts [16]. This could be done in particular by using existing ontologies (or new ones) which facilitate the extraction of data semantics.

Predictive Analytics can be defined as the incorporation of mathematical and sophisticated statistical techniques in building a model that can be used to forecast the future and discover predictive patterns from historical datasets. [17]. Predictive models can be classified based on the nature of the intended task or by the embedded architecture of the predictive model. In other words, predictive models can be categories as regression models (continuous output) or classification model (nominal or binary output). Also, based on the architecture of the embedding model, the predictive models can also be grouped based on their technical implementation. These include Linear models, Decision trees (also known as Classification and Regression Trees or CART), Neural networks, Support vector machines, Cluster models, etc. [18]. Further details on the algorithms and techniques used for predictive analytics will be discussed in the following section.

In this section we provide a conceptual framework solution for predictive analytics in SBD which addresses Big data features. Figure 1 shows the proposed framework which comprises certain stages that commonly followed when carrying out predictive analytics in SBD. These stages can be briefed as follows:

**Stage 1- SBD Generation:** At this stage, heterogeneous types of data are generated from different sources which form the *variety* dimension of Big data. These data islands include data exists as mainly in an unstructured format (such as social media streams, web blogs, news agencies, reviews etc.).

**Stage 2- SBD Acquisition:** this stage comprises gathering data from the aforementioned data sources into unified data repository. Data can be collected using various Application Programing Interfaces (APIs) provided by data sources to facilitate data gathering. For example, The Twitter (API) offers a predefined way to access and retrieve tweets [19] which can be utilized to extract batches of tweets in a timely fashion. The raw extracted data will pass through a pre-processing phase. This phase will address the data *veracity* via data correctness. This phase includes:

(i) Data cleansing: data at this stage may include many errors, incomplete, inaccurate, meaningless data, irrelevant data, redundant data, etc. Thus, data cleansing will remove dirty and corrupted data and ensure data consistency;

(ii) Temporary data storage where data is grouped and stored in a temporary location before it can be departed in the distributed data store;

(iii) Data integration: through data reformatting to fit with the predefined data structure model that is designed basically based on the tweet's metadata. Besides the unstructured data collected from Web resources, data can also be obtained from structured data islands (such as data mainly coming from day-to-day operational databases and structured external data sources). Structured data can be consolidated with the pre-processed social data to benefit Business Intelligence applications.

**Stage 3- Data Storage**: handling large scale data, such as social data, necessitates the need to develop a distributed data processing solution to facilitate data storage and analysis. The Big data storage provides distributed and parallel data processing infrastructure based on the Hadoop/MapReduce platform for Big data to store and manage such massive amounts of data. The temporal-temporary data dumps its contents to this Non-Volatile distributed environment after the integration process. This dump will be assigned a timestamp to differentiate it from previous batches. These data silos can be then used for conducting further analysis, also it can be integrated with data warehouses.

**Stage 4- Data Analytics**: This is the focal area of SBD predictive framework. In this stage, datasets will be collected from the distributed environment using NoSQL Language. It is important to recall the identified problem that might describe a business need. Then in order to translate this problem to a proper predictive analytical solution, this requires incorporating the right datasets. For example, the problem pertaining solely on evaluating credibility of users in a certain OSN platform should entail obtaining cleansed datasets collected from users of such platform and also other datasets embody information and metadata about the same users, but neglect datasets attained form other data resources where information about those users are absent.

Constructing an efficient predictive engineering model requires to assure data is well prepared for this task. Also, it involves selecting the best features that can be used along with the best machine learning model, this is followed by the evaluation protocol to measure the performance of the predictive model. In particular, the following steps can be commonly followed in an iterative manner for an efficient predictive modelling:

(i) *Data Preparation and Normalisation*: this step has been partially attained in the pre-processing phase where data is cleansed and integrated for further data analysis. Yet, at this phase, commonly normalization techniques can be also applied such as: data scaling (transforming floating point values to a certain standard range – commonly zero to one); clipping (involves capping values of potential features that contain outliers to be under (or above) a certain threshold); Logarithmic scaling(commonly used to alleviating variance in datasets by squeezing wide range values).

(ii) *Feature Engineering and Selection*: this step is crucial in the predictive modeling as it involves two vital sub steps, namely: Feature Engineering and Feature Selection. Feature Engineering (or extraction) aims to infer and identify the most important features from datasets that can be used to address a certain domain-based problem. It requires applying domain knowledge on the dataset, thereby extracting all possible features that can be potentially incorporated in a predictive modelling task. Feature Selection follows Feature Engineering which aims to choose the optimal features amongst the candidate ones. Feature selection is critical and affects the performance of the predictive model. Therefore, adequate feature selection techniques should be carried out to infer the best features. These techniques involve Pearson Correlation, Chi-Squared, Tree-based to name a few.

(iii) *Model Selection*: one of the challenges in predictive modeling process is to choose the best machine learning model for a certain examined problem. In fact, this challenge can be also forked into two sub-challenges, namely: selecting model amongst a set of candidate models and also selecting the best hyperparameters settings of the same selected model [20]. Yet, the question is what are the criteria that should be considered in such selection process? This decision is made by reaching a compromise between various factors that play a vital role in this process such as: benchmark comparison amongst state-of-the-art models, model performance within available time, skills, and resources, etc.

Model selection techniques can be commonly categorized into: Probabilistic measures (evaluate the candidate model considering it complexity and performance on the training dataset), and Resampling techniques (using approaches such as cross-validation). Hyperparameter selection is another category of model selection. This category comprises certain techniques that are used to search for the best hyperparameters settings, such as: Grid Search (searching the best hyperparameters using a grid of specified parameter values); random search (seeking the model with the optimal hyperparameters in a valid search space using a specified number of trials); evolutionary algorithms (finding optimal hyperparameters using heuristic-based approaches); and Bayesian optimization (a probabilistic model used for hyperparameters tuning).

(iv) *Model Evaluation*: as discussed previously, the optimal predictive model is selected using various selection techniques, in which the performance can be measured on the training dataset. Yet, to study how the model generalizes on unseen dataset is another important consideration. This is a decisive step in the predictive modeling framework as it creates venues for continuous improvements on the implemented approach. In fact, building a robust predictive model is not a trivial task; it entails carrying out solid measures to quantify its effectiveness and accuracy. There are several metrics that can be used to evaluate the performance of a machine learning model. These measures can be mainly classified to two categories: Regression metrics and Classification metrics. Regression metrics include measures such: RMSE (Root Mean Square Error), MAE (Mean Absolute Error), R Squared ($R^2$) and Adjusted R Squared. Classification metrics embody certain well-known metrics such as: Precision, Recall, F1 Score Accuracy, ROC, AUC, Log-Loss etc.

**Stage 4- Integration and Visualization**: Business Intelligence applications are designed mainly to process and manipulate structured data stored in data warehouses. Therefore, another significant task of predictive modeling framework is to articulate structure from unstructured input data. In this context, the data structuring process incorporates semantic analysis so as to provide a better understanding of the schema structure of data warehouses. Thus, the next process will migrate these data with the data coming from the ETL (Extract, Transform and Load) process to produce integrated, meaningful, and trustworthy structured data to load into the integrated data warehouse. This integration of traditional business intelligence and social business intelligence will enhance the quality of reports extracted by BI tools, thus achieving an additional objective of the overall framework.

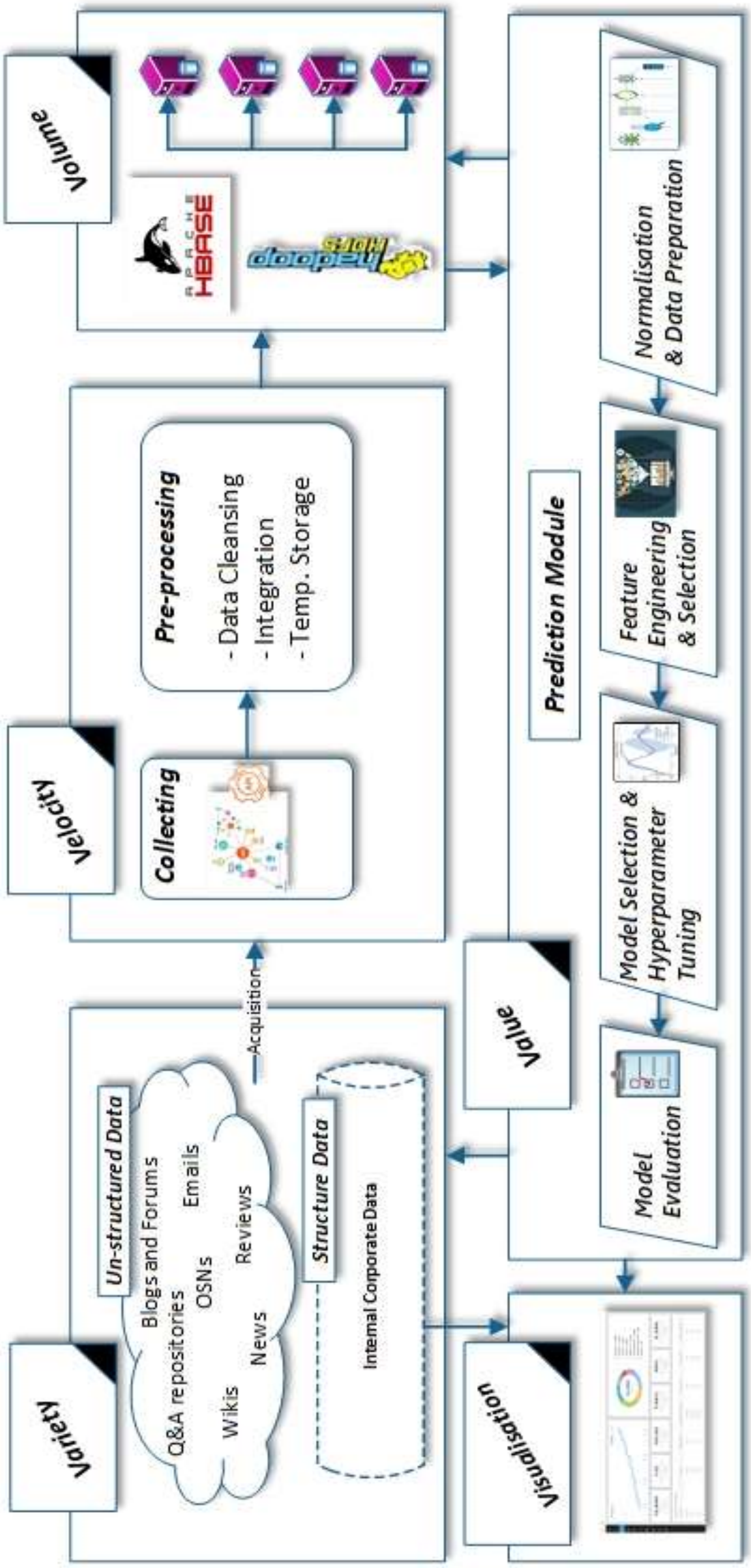

Figure 1: Predictive Analytics Framework for Social Big Data

## 5.3 Machine learning algorithms for SBD predictive analytics

In this section, we discuss several commonly used machine learning algorithms to develop a predictive model, which predicts the new social situation, $\hat{y}$ , when the prediction features, $\bar{x} = (x_1, x_2, ..., x_n)$, are given. The predictive model can be developed when a dataset which relating the prediction features and past social situation, $y$ , is available, where $n$ is the number of prediction features. We discuss the following machine algorithms namely logistic regression (LR), generalized linear model, naïve bayes (NB), tree based classification, random forest, gradient boosted tree and deep learning:

### 5.3.1 Logistic Regression

Logistic regression is a statistical method which has commonly been using for binary or diagnosis classification [21]. The approach predicts whether the predicted new social situation, $\hat{y}$ , is between zero to one when the prediction features, $\bar{x}$ , are given. In this regression approach, we assumed that the new social situation is positive when $\hat{y}$ is larger than 0.5. Otherwise, the new social situation is negative when $\hat{y}$ is smaller than 0.5. When $\hat{y}$ is close to one, the likelihood of positive new social situation is high. When $\hat{y}$ is close to zero, the likelihood of negative new social situation is high. The logistic regression can be formulated as,

$$F^{Logistic}(\bar{x}) = \hat{y} = P(\hat{y} = 1 | \bar{x}) = \frac{1}{1 + \exp\left(-\left(b_0 + \sum_{i=1}^{n} b_i \cdot x_i\right)\right)} \quad (5.1)$$

where $\bar{b} = (b_1, b_2, ..., b_n)$ denotes the set of logistic parameters. $\bar{b}$ is optimized by maximizing the likelihoods for all samples. Newton's iteration method can be used to determine the optimal $\bar{b}$ which attempts to fit the predictions and the real social situation. $\bar{b}$ is iteratively adjusted using the gradient between $\hat{y}$ and $\bar{x}$ . The optimization process is terminated when the difference between $\hat{y}$ and the real observation cannot be deceased.

### 5.3.2 Generalized linear regression

Another approach namely generalized linear regression [21] in (5.2) can be used to predict the new social situation.

$$F^{generalized}\left(\bar{\beta}\middle|y,\bar{x}\right)=\hat{y}=\sum_{i=1}^{N_D}\left(y_i-\left(b_0+\bar{x}\cdot\bar{\beta}^t\right)\right)+\lambda\left(\frac{1-\alpha}{2}\left\|\bar{\beta}\right\|_2+\alpha\left\|\bar{\beta}\right\|_1\right) \quad (5.2)$$

where $N_D$ is the number past samples.

The approach is another version of logistic regression which minimizes the differences between the predicted social situations $\hat{y}$ and the real observations. (5.2) is involved with the penalty term which is the second term and is used to avoid overfit the real observations. The penalty term consists of norm-1 and norm-2 of all coefficients $\bar{\beta}=\left(\beta_0,\beta_1,\beta_2,...,\beta_n\right)$ in order to perform the regulaziation of the fitting. In (5.2), the regularization parameter is denoted as λ. Large $\beta_i$ with either $i=0,1,2,...,$ or $n$ is restricted by the norm-2 and a large $\alpha$. All $\beta_i$ are equally restricted if $\alpha$ is a small value. $\bar{\beta}$ is determined by minimizing (5.2) when the samples of the past social situations and the corresponding prediction features are given.

### 5.3.3 Naïve bayes (NB)

Naive Bayes [22] is an approach of classifier which is generally more capable to model a complicated or nonlinear social situation, compared to the statistical regressions. The approach has been applied effectively on text/opinion mining, spam detection, and decision recommendations. An advantage of Naïve Bayes Classifier is that the computational cost of developing the classifier is generally lower and the generalization capability is generatlly higher, compared to the other machine learning approaches such as the deep neural networks. The approach is developed based on the Bayes' Theorem, where the prediction features are mutually independent. The prediction model can be developed in a short amount of training when huge amount of data is given for modelling. In the Naive Bayes model, the posterior probability, $P\left(c\mid\bar{x}\right)$, is defined in equation (5.3), where $c$ is either positive or negative social situations. $P\left(c\mid\bar{x}\right)$ is the likelihood of whether $c$ is positive or negative. Given that the domain is $c$, $P\left(x_i\middle|c\right)$ is the probability that the feature is $x_i$. Based on the Bayes' Theorem, $P\left(c\right)$ is the probability that the social situation is $c$.

$$P\left(c\middle|\bar{x}\right)=P\left(x_1\middle|c\right)\times P\left(x_2\middle|c\right)\times...\times P\left(x_n\middle|c\right)\times P\left(c\right) \quad (5.3)$$

### 5.3.4 Decision tree based classification

The decision tree based classification is an expression of a recursive partition of the social situation. The decision tree is similar to the flow chart. The top of the tree is the root node which holds the final prediction namely the new social situation, $c$. The tree bottom consists of terminal node namely leaves and each leave is represented by $x_i$ with $i=1,2,...,n$. Each leaf votes what the final prediction is. The tree branches generate the outcomes which are interconnected with all $x_i$. The final prediction is the overall average of the branches which have the root of all $x_i$. In other words, the final prediction of the decision tree is the majority of the leave contents. When developing the decision tree, the branches of the tree are expanded repeatedly by including more new nodes. The development is terminated until the stopping criteria of the training is satisfied. The stopping criteria can be predefined as a certain number of iterations or a certain prediction accuracy. The decision trees can be generated by the three commonly used approach namely C4.5 [23], random forest [24] and gradient boosting [25].

Decision trees are more transparent and easier to explain the relationship between the prediction features and the social situation, compared to logistic regression, generalized linear regression, and naïve bayes. They nevertheless attempt to model the relationship between the prediction features and the social situation. Caruana et al. [26] stated that the classification performance of decision tree is better than the logistic regression and support vector machine when solving the 11 classification benchmark problems; ten classification metrics are used for the evaluations.

### 5.3.5 Random forest

The approach of random forests is a new version of decision tree. It attempts to incorporate several decision trees. It is a hybrid decision tree which averages the outcomes of several decision trees, where each decision tree is generated based on different training subset of the whole training set. The random forest model namely is formulated by equation (10).

$$F^{Rand}\left(\bar{x}\right)=\frac{1}{N_{DT}}\sum_{i=1}^{N_{DT}}F_i^{DT}\left(\bar{x}\right) \qquad (5.4)$$

where $F^{Rand}\left(\bar{x}\right)$ is the overall average of all $F_i^{DT}\left(\bar{x}\right)$ with $i=1,2,...,N_{DT}$. All $F_i^{DT}\left(\bar{x}\right)$ are developed by different subset of training set. $F_i^{DT}\left(\bar{x}\right)$ can be developed by either C4.5 [23], random forest [24] or gradient boosting [25]. The approach of random forests attempts to overcome the limitation of the single decision trees, which can be overtrained when running indefinitely.

### 5.3.6 Gradient boosted tree

The hybrid tree in (5.4) can be modified as (5.5) of which the weights of each decision tree are unequal; each decision tree is weighted with a different constant, $\alpha_i$ with $i=1,2,...,N_{DT}$.

$$F^{GB}\left(\bar{x}\right)=\frac{1}{N_{DT}}\sum_{i=1}^{N_{DT}}\alpha_i F_i^{DT}\left(\bar{x}\right) \qquad (5.5)$$

where $\sum_{i=1}^{N_{DT}} \alpha_i = 1$ and all $\alpha_i$ ‘s are optimized by minimizing the mean square error between the real observations and the predictions of $F^{GB}(\bar{x})$. The optimization can be formulated as a lagrange multiplier in order to determine the optimal $\alpha_i$ ‘s [27]. The generalization capabilities of $F^{GB}(\bar{x})$ in (5.5) are generally better than those of $F^{Rand}(\bar{x})$ in (5.4). In $F^{GB}(\bar{x})$, a larger $\alpha_i$ is multiplied with a $F_i^{DT}(\bar{x})$ which has better generalized capability.

### 5.3.7 Deep Learning

The original configuration of deep learning (DL) is a multi-layer feed-forward artificial neural network which has more than two hidden layers. The prediction features, $\bar{x}$, are the network inputs and the network output is the new social situation, $y$. In the network, the neurons are interconnected and each neuron is stimulated with a activation function. Tanh, rectifier or maxout functions can be the activation function, which models the nonlinear relation between prediction features and the new social situation.

Neurons are linked by the weights; each neuron output is multiplied by the weight and the neurons outputs are feed into other neuron inputs. Based on this complex network structure, very complicated or nonlinear relation can be modelled. The weights are generally determined by the gradient descent algorithm such as the back-propagation. A network with high generalization capability can be developed by the advanced technologies such as regularization, dynamic learning rate, neuron dropout. Generally the generalization capability of the deep learning are better than the classical classification methods such as the statistical regression, decision tree.

## 5.4 Predictive analytics applications, tools and APIs for SBD

Machine learning applications enable real-time predictions by leveraging high quality and well-proven learning algorithms. Based on the current dominant position and high impact on business in several use cases, predictive analytics, in particular, enhances the decision-making process and provides valuable insights on large-scale datasets. In the context of SBD, businesses have benefited from the prevalence of social media services as these enable them to establish interactive-based dialogues with their customers [28]. These dialogues which appear in companies’ social pages permit customers to express their views freely and without restrictions on products and services provided by those companies, using comments or replies to either praise these products or services, or to point out their shortcomings. This indeed provides an opportunity for the business firms to study and respond to these opinions, thereby enhancing its customer service, which greatly extends their customer knowledge, customer acquisition, and customer retention [29, 30]. Next section presents a selected set of predictive analytical applications that benefit from the social data.

### 5.4.1 Applications on incorporating predictive analytics for SBD

#### 5.4.1.1 User modelling and personalization

User modelling is the process of framing user's characteristics collected from various resources into a unified user profile that demonstrate user behaviour, interests, knowledge and expertise in countless applications [31, 32]. Social data provides tremendous amount of user-specific data and metadata that can be used in personalisation practices in dissimilar domains. For example, in e-learning environment, observed social data pertaining to personal information, learning activities, preferences, sentiments, comments, and learning style can be collected and used to infer user features [33]. These features that demonstrate social learner behaviour can be then used to carry out predictive analytics on user modelling, and predict for example student performance [34], identify patterns of students' learning weaknesses and strengths, provide tailored recommended educational material and courses, etc. Another important use of personalisation and user modelling is incorporating social data collected from online social media services to benefit health domain. For example, collected social data of users can lead to detecting psychological patterns, identifying users' mental health status [35], predicting depression [36] and personal treats [37].

#### 5.4.1.2 Spam and social influence prediction

Social media services have been increasingly targeted to spread spam and other malicious activities [38]. Spammers misuse the social media services features and tools, sending annoying messages to legitimate users, publishing contents that include malicious links, and hijacking popular topics [39]. They post contents on various topics, and they duplicate posts [38]. Further, to propagate their vicious activities, spammers abuse other social media services features such as hashtags, and mention other users and link-shortening services [40]. On the other hand, social media services are a rich platform through which users can express their opinions and share their views, experiences and knowledge of numerous topics. Hence, discovering users' influence has been motivated by its significance in a broad range of applications such as personalized recommendation systems [41], opinion analysis [42], expertise retrieval [43], and computational advertising [44]. Therefore, examining users' trustworthiness over social media plays an important role in determining whether the information being offered can be trusted. Since much of this information has been contributed by users with limited or no credible history, the task of detecting content credibility is challenging. Hence, predicting credibility in the context of spam and user influence is crucial and requires a deep scrutiny to multimodal data generated by users of social environment [2, 4, 12, 45, 46].

#### 5.4.1.3 Content segmentation and classification

Segmentation is the first step towards effective marketing, and is intended to classify customers according to their interests, needs, geographical locations, purchasing habits, lifestyle, financial status and level of brand interaction [47]. If companies succeed in building predictive analytics that obtained effective clusters of customers and determining the basic criteria for each cluster in making their buying decisions, companies will be able to establish goals and take appropriate actions to achieve them. For example, companies can identify the most optimal products/services captured for each segment of customers. This fine-grained analysis can maximise customer satisfaction as companies can then design and manufacture not only one standard product, but several segment-oriented products. Further, since users of online social services are keen to establish strong relationships with others; they search for and connect with relevant content or users. Hence, to open dialogues between like-minded people so as to share opinions, and life experiences, the first step is to obtain a holistc view of user interest automatically. This is achieved through the analysis of users' content and determining all behavioural aspects of users. This allows the segmenting and searching of users according to their domains of interest [48]. For example, supervised machine learning techniques are used to perform domain-based classification task for the semantically-enriched temporally-segmented social textual content [3].

#### 5.4.1.4 Customer engagement:

Online social services offer widely significant implications for a variety of business-related applications such as the VoC/VoM, recommendation systems, the discovery of domain-based influencers, and opinion mining through tracking and simulation. In particular, the accurate understanding of the domains of interest extracted from user social data strengthens the engagement between businesses and their current and prospective customers. This contributes to an accurate analysis of indirect customer feedback that comprises social listening to customer reviews and opinions to improve brand loyalty, communication, customer service, customer care interactions, etc. Hence, "many marketing researchers believe that social media analytics presents a unique opportunity for businesses to treat the market as a 'conversation' between businesses and customers" [28]. Therefore, efforts have been conducted to incorporate social media serives to boost customer to business engagement in tourism [49], brand loyalty programs [50], social media marketing [51] to various other applications. Further, predictive analytics advances these endevours by offering a room to identify the optimal dimensoins to model customer engagement. For example, machine leanring applications can be used to dynamically personalize the customer engagement experience thereby identifying what customers value in real time [50].

### 5.4.2 Predictive analytics tools and APIs

The ever increase in incorporating Artificial Intelligence technologies to tackle large-scale problems has urged tech business firms to offer a wide range of machine learning-based platforms. These platforms are particularly crafted to embed cutting edge algorithms and sophisticated mathematical and statistical modules yet friendly GUIs and easy access APIs. The Gartner magic quadrant for data science and machine learning platforms indicate a graphical competitive positioning of four categories of distinguished technology vendors for data science. These categories or quadrants are: *Challengers, Leaders, Niche Players and Visionaries*[1]. As depicted in Figure 2 sixteen different technology providers have been divided into the designated four categories.

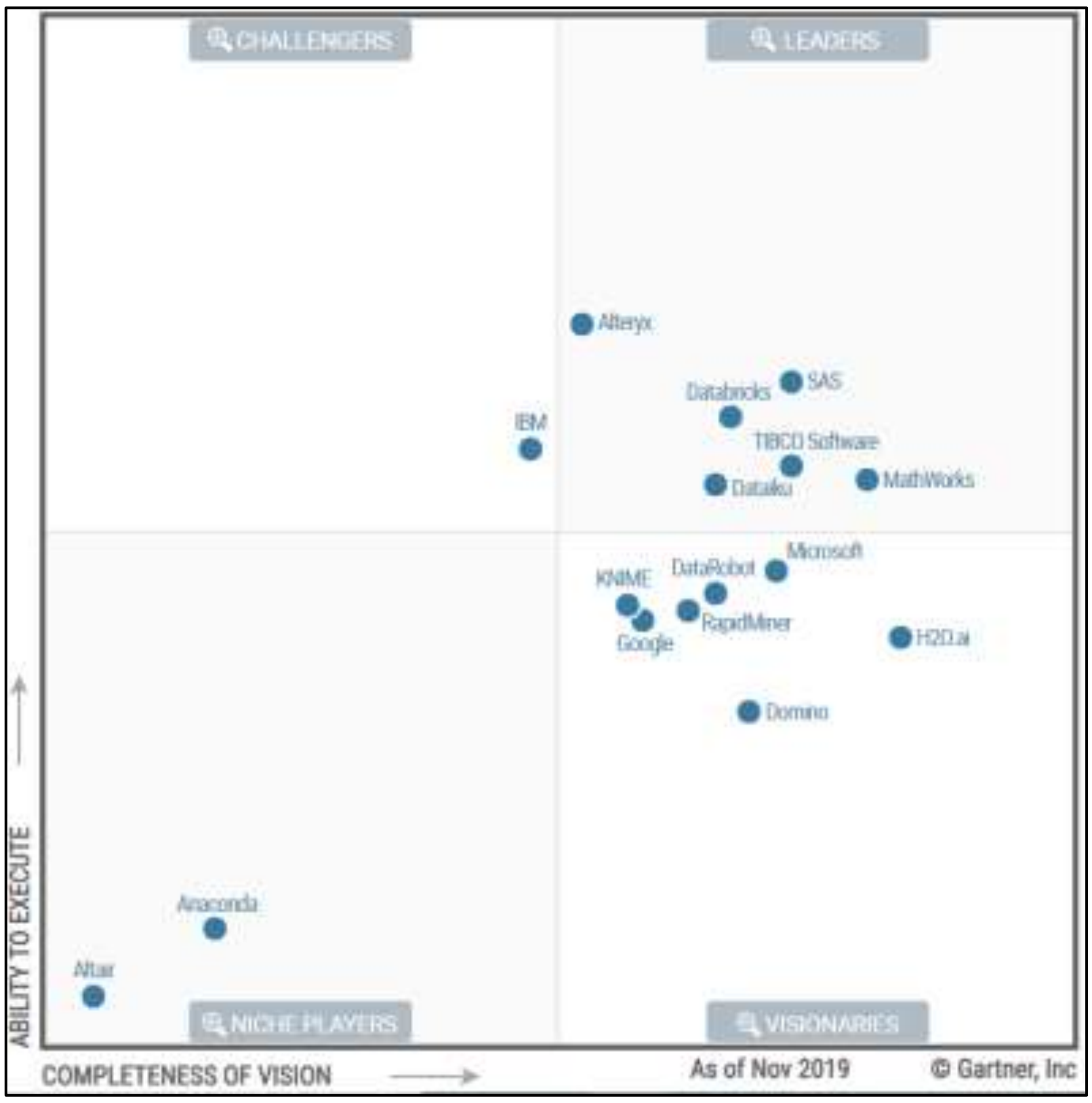

Figure 2: Gartner magic quadrant for data science and machine learning platforms (Feb 2020) [2]

---

[1] Reader can refer to https://www.gartner.com/en/research/methodologies/magic-quadrants-research to obtain further discussion on these four quadrants.

[2] https://www.gartner.com/document/3980855?ref=solrAll&refval=250693063

The following is a brief overview to a selected set of platforms and tools that are produced by vendors indicated in Gartner magic quadrant. These products can be accommodated for conducting predictive analytics on SBD.

#### 5.4.2.1 RapidMiner Studio™

RapidMiner Studio™ is an open-source on-premise predictive analytical software product that is designed and developed by the visionary company RapidMiner [52]. RapidMiner studio offers distinctive features which facilitate carrying out predictive analytical modelling by novice citizen data scientists using friendly interface. It enables also expert data scientists to conduct optimisation and hyperparameter tuning on solutions being implemented. Internal modules of RapidMiner studio can be integrated and further extended with various components and technologies provided into their marketplace. These extensions include: Textual processing tools, Python scripting, Wordnet extension, Weka extension, to name a few. Further, RapidMiner studio provides capacity to establish end-to-end machine learning development life cycle pipeline that can be initiated with cleansing raw collected data and concludes with a deployed trained and evaluated product. RapidMiner has also launched RapidMiner Go™, an online cloud-based automated machine learning platform, that can be used as a collaborative environment for data analysts to build and evaluate predictive applications.

#### 5.4.2.2 SAS® Visual Data Mining and Machine Learning

SAS® Visual Data Mining and Machine Learning (VDMML) is core analytical model designed and developed by a leader firm SAS®. The architecture of VDMML embodies modular components each indicate a certain atomic analytical activity, such as variables selection, module construction, output generation, etc. These comprised activities can be accessed through various interfaces and RESTful APIs services [53]. VDMML ecosystem is solidified with several distinctive features that give the application a superiority among other alternatives. This is evident as VDMML provides capacity to conduct analytics using automated feature engineering and modelling and thereby generate insights with minimal learning time and efforts. Also, the internal architecture of SAS® VDMML is designed to handle large-scale datasets. This is through enabling distributed processing across a cluster of nodes as well as efficient memory utilisation and computation [54].

#### 5.4.2.3 TIBCO® Data Science

With being indicated as a leader by Gartner and Forrester research [55], TIBCO Software has been continuously evolving and positioning itself in midst of the big players of data science and predictive analytics. Amongst featured products developed by TIBCO, Data Science software is tailored to allow data analytics practitioners, experts, team members and citizens to carry out predictive analytical experiments using pre-built template, visualisation technologies, and sophisticated machine learning algorithms and modules [56, 57] . With wide variety of internal frameworks, open source tools and technologies, TIBCO Data Science offer users a comprehensive set of toolkits to enable them constructing complex solutions to tackle dissimilar real-life problems. For example, TIBCO Data Science platform is a good choice for undertaking social media analysis, as it provides mechanisms for data monitoring, textual analytics, time-series forecasting, to name a few [58].

#### 5.4.2.4 H2O Driverless AI

H2O Driverless AI is an open-source data science framework developed by a visionary company H2O.ai. H2O Driverless AI has maintained a good reputation due to its built-in advanced algorithms that include deep learning, boosting, and bagging ensembles [59]. H2O has also succeeded to make its AI framework elastic in essence that most of the internal activities can be undertaken automatically without much of user involvement. These activities include feature selection and engineering, model selection and validation, and model deployment [60]. Further, H2O provides convenient access APIs and interfaces for many scripting languages and it has proven ability to accommodate billions of records efficiently using in-memory compression [61]. In the context of SBD, H2O Driverless AI has been used as an effective tool for example to carry out social data wrangling, social sentiment analysis and social news mining [62].

## 5.5 Case study on social politics domain

In this section we will follow the framework depicted in Section 2 by conducting an experimental analysis on politics domain. In particular, we will carry out a domain-based classification to a select set of users collected from Twitter. These users are divided into two categories, namely users who are interested in politics domain and users do not show an interest in this designated domain.

### 5.5.1 Data generation and acquisition

Twitter has provided a rich dataset of over 500 million tweets daily which is around 200 billion tweets a year [63]. Twitter mining is an emerging research field falling under the umbrella of data mining and machine learning. Twitter is the chosen subject of this research because: (1) Twitter is a fertile medium for researchers in diverse disciplines, leveraging the vast volume of content; (2) Twitter facilitates data collection by providing easy access APIs to The Twitter-sphere; (3) due to the economy and the ambiguity and brevity of a tweet’s content, it is challenging to determine the accurate domain(s) to which the user’s tweet is referring.

For proof of concept, this study is limited to an on/off domain classification to users of Twitter and the political domain has been selected for the following reasons: (1) Twitter has been intensively incorporated as an important arena by politicians to express and defend their policies, to practice electoral propaganda and to communicate with their supporters [64], (2) Twitter has raised considerable controversy regarding its usage as a platform to attack political opponents [65], (3) Twitter is characterised by its growing social base to include broad political, social groups leveraged by ease of use, free access, and deregulated nature [66], (4) the amount of the political discourse in social content is overwhelming; over one-third of OSNS's users believe that they are worn-out by the quantity of the political content they encounter [67]. Such an abundance of data facilitates data aggregation and improves the outcome of the data analysis. For future work, this study aims to develop a multi-domain-based classification, leveraged by domain ontologies, semantic technologies and linked open data. Hence, besides the political domain, an analysis of other domains of interest may be further investigated in the future.

The dataset used for this study has been collected using Twitter's "User_timeline[3]" API method. This mechanism allows access to and retrieval of public users' content and metadata. The collection of the users' content was accomplished in two stages: (1) by collecting historical user content (up to "3,200" most recent tweets[4]). This dataset will be used to predict the user's interest in the political domain in general, and (2) by collecting the new content of those users whose historical tweets were obtained in the first phase. As will be described later, the dataset of the first stage is used to predict the user's interest in politics at the user level, i.e. to establish an understanding of the user's interest in the political domain based on the user's past content.

### 5.5.2 Dataset pre-processing

The veracity of data refers to the certainty, faultlessness and truthfulness of data [68]. Although availably, reliability and security of data's nascence and storage are significant, these factors do not guarantee data correctness and consistency. Appropriate data cleansing and integration techniques should be incorporated to ensure certainty of data. The data collected for the user's content, and historical and new tweets, are pre-processed by data quality enhancement and data cleansing techniques which are discussed below:

**Data cleansing** of user content is conducted by using the following techniques: (1) all redundant content (i.e. same dataset collected more than once) such as tweets or user data is eliminated with their metadata; (2) removing stop words; (3) removing URLs; (4) decoding all HTML entities to their applicable characters; (5) removing punctuation marks, correcting encoding format, etc.

**Quality enhancements,** the list of Twitter handles (a.k.a. Twitter user/screen name such as @example), which are indicated in the user's tweets, is collected and replaced with the user's corresponding names. This is achieved through the Application Programming Interface (API) of Twitter's "lookup[5]". These handles are normally neglected or deleted when mining user's tweets. However, these handles are important because they are used by Twitter users to mention other Twitter users in their tweets, replies or re-tweets. Hence, it is essential to identify and ascertain the actual names of those users. This assists in the process of domain extraction. For example, a user shows an interest in the political domain if she/he commonly indicates handles linked to politicians or political parties, in addition to publishing other politics-related content.

---

3 https://dev.twitter.com/rest/reference/get/statuses/user_timeline.

4 This threshold is set by Twitter™ as the maximum number of recent tweets the twitter API is allowed to retrieve.

5 https://dev.twitter.com/rest/reference/get/statuses/lookup.

### 5.5.3 Feature Engineering and Selection

The pre-processed dataset is passed to the prediction module. This study aims to establish a fundamental ground for efficiently detecting the domain of interest of Twitter users, which will significantly contribute to a better understanding of the domain(s) of future users' tweets. As a proof of concept, the proposed system is validated by an application on the Political domain, where the proposed system attempts to detect whether the domain of interest of a user is or is not politics related. This validation is based primarily on former knowledge about a user's political interests obtained by analysing the user's historical content. To do so, the following politics-domain knowledge inference approach is designed to extract the semantics of a user's tweets, thereby uncovering the user's domain of interest.

#### 5.5.3.1 Political domain Knowledge Inference

In the feature extraction module, domain knowledge inference is the main process used to extract user and tweet features from pre-processed datasets. For proof of concept, the study focuses on the political domain, using politics ontology, WordNet, and ontology interoperability to infer politics knowledge.

**Politics Ontology and WordNet®:** The political domain refers to the knowledge captured in politics ontology along with its knowledge base. BBC defines politics ontology as "an ontology which describes a model for politics, specifically regarding local government and elections" [69]. The BBC Politics ontology conceptualises a politics model especially for the UK government and elections. It was originally designed to cope with UK local government and European Elections in May 2014. This study applies the BBC Politics ontology to Australian politics by further extending politics concepts. BBC Politics ontology and its extension ontology harnessed in this chapter are depicted in xxx and yyy respectively. Furthermore, this study uses WordNet[6], which is a lexical dictionary used to construct relations between terms of synonymies. Synonyms (or synsets) are a set of interrelated terms or phrases which indicate the same semantic concept, such as the words "elections, public opinion poll, opinion poll, and ballot". All the synsets of the political concepts captured in politics ontology depicted in yyy are examined, and only the synonyms applicable to the political context are captured.

**Ontology Interoperability** :The interlinking with other relevant entities defined in other datasets supports interoperability [45, 46, 70, 71]. The approach taken in this study addresses information interoperability by focusing on the equivalence links that direct the URI to refer to the same resource or entity. The politics ontology supports the equivalence links between the ontology components and the tweet data. The resources and entities are linked through the owl#sameAs relation. This implies that the subject URI and object URI resources are the same, and hence the data can be further explored.

---

6 https://wordnet.princeton.edu/

In the interlinking process, IBM Watson NLU ™ is incorporated as a one-stop shop, leveraging access to a wide variety of linked data resources[7] through providing easy access APIs. These resources include but are not limited to: different vocabularies such as Upper Mapping and Binding Exchange Layer (UMBEL), Freebase (which is a community-curated database for well-known people, places, and things), YAGO high-quality knowledge base, and DBpedia knowledge base, etc. These resources are used to help extend the knowledge base of the politics ontology by identifying (non-)Australian politicians and (non-)Australian political parties from users' tweets. For example, at this stage, "99,812" instances of "2009" politicians, and "48,704" instances of "59" political parties are captured in the politics ontology.

#### 5.5.3.2 User Features

The political interest of users is primarily measured by two main proposed factors: continuity and knowledgeability. Continuity refers to the frequent interest of a user in a certain domain. In other words, the user demonstrates an interest in the political domain by tweeting or retweeting content in this domain over a relatively long period. Continuity is measured by counting the number of political entities identified from the user' tweets in each period (such as every month, quarter, etc.). Knowledgeability (or Speciality) refers to the user's close acquaintance with the political domain and also refers to the user's dedicated pursuit of the political domain through a commitment such as work or study. Knowledgeability is measured by accumulating the distinct number of political entities annotated from the user's tweet, and the user's profile description [72-75]. Table 1 shows the list of features used to classify whether the user's interest is **on-topic** or **off-topic**. **On-topic** refers to when the user demonstrates a continuous interest in the political domain. **Off-topic** users are those whose Twitter content shows their non-interest in the political domain.

Table 1: A List of User's Features

| No | Features | Description |
|---|---|---|
| 1 | no_tweets, $x_1$ | The total count of users' historical collected tweets up to 3,200 tweets. |
| 2 | unq_pol_entities, $x_2$ | Total count of distinct/unique political entities extracted from all user's tweets |

7 http://www.alchemyapi.com/products/alchemylanguage/linked-data.

| | | |
|---|---|---|
| 3 | pol_entities_pre_QW_YYYY, $x_3$ | Count of political entities annotated from the tweets posted before quarter 'W' of the year 'YYYY' |
| 4 | pol_entities_QW_YYYY, $x_4$ | Count of political entities annotated from the tweets posted in quarter 'W' of the year 'YYYY' |
| 5 | pol_entities_QX_YYYY, $x_5$ | Count of political entities annotated from the tweets posted in quarter 'X' of the year 'YYYY' |
| 6 | pol_entities_QY_YYYY, $x_6$ | Count of political entities annotated from the tweets posted in quarter 'Y' of the year 'YYYY' |
| 7 | pol_entities_QZ_YYYY, $x_7$ | Count of political entities annotated from the tweets posted in quarter 'Z' of the year 'YYYY' |
| 8 | profile_pol_entities, $x_8$ | Count of political entities annotated from user's profile description |
| 9 | verified(Authentication Status), $x_9$ | Authentication flag used for accounts of public interest (for example, politicians) |
| 10 | fav_count, $x_{10}$ | Total count of likes the user has |
| 11 | replies_count, $x_{11}$ | Total count of replies the user has |
| 12 | retweet_counts, $x_{12}$ | Total retweets of the user's tweets |
| 13 | followers_count, $x_{13}$ | Number of user's followers (who follow the user) |
| 14 | friends_count, $x_{14}$ | Number of user's friends (followed by the user) |

The features $x_2$ to $x_8$ as depicted in Table 1 are selected to primarily focus on users' ongoing interest in and knowledge about the political domain by extracting the political entities from their tweets and by leveraging the knowledge-inference tools and repositories explained in section 2. In particular, features $x_2$ to $x_8$ are proposed to address the political knowledgeability of users. Moreover, features $x_3$ to $x_7$ address the continuing interest of users in the political domain. Features $x_1$ and $x_9$ are added to support the aforementioned features and will be discussed later in this study.

Unq_pol_entities ($x_2$), listed in Table 1, refers to the number of distinct political entities extracted from the history of a user's tweets. Profile_pol_entities ($x_8$) represents the number of all political concepts that are identified in the users' profile description on their Twitter accounts. The former feature represents the diversity of the political concepts embodied in the users' tweets, and the latter feature, $x_8$, is used to examine the explicit indication of the user's interest in the political domain, particularly if the user works in this domain. This is usually clearly indicated in their profile description.

The list of all political entities is counted periodically. The political entities extracted from the user content for each period is used to scrutinise political interest temporally rather than scrutinising the tweets as a whole. Therefore, the collected historical tweets are divided into five groups, $x_3$ to $x_7$. Four groups, $x_4$ to $x_7$, indicate the four sequential and recent quarters (W, X, Y and Z), where ‘Z’ is the most recent quarter and one group, $x_3$, indicates the rest of the tweets posted before the ‘W’ quarter. This mechanism is proposed because the user’s interest(s) may change, and their knowledge may evolve. Hence, it is more efficient to examine the user’s domain(s) of interest based on current and recent behaviours from the four-time groups. Furthermore, some users only show a particular interest in the political domain when popular political events are taking place. For example, a users’ involvement in conversations during election campaigns does not necessarily indicate an interest in the political domain generally, as the election is a trending topic only, on which users with dissimilar interests share their thoughts, and/or anticipations about the potential candidates.

The remaining seven features listed in Table 1 are the no_tweets, verified status, favorites count, replies count, retweet counts, followers count, and friends count. The no_tweets, $x_1$, relates to the number of collected historical tweets. This feature is important as a means of addressing the ratio between the number of political concepts accumulated for features $x_2$ to $x_8$ and the total number of tweets. For example, two users might archive the same number of distinct political concepts, although the number of tweets differs for each user. The verified feature, $x_9$, is the authenticated flag (i.e. blue verified badge ). Twitter may set this flag to ‘1’ for users of public interest. Twitter currently offers this feature to help users find influential and high quality accounts in several domains[8]. Features $x_{10}$ to $x_{14}$ indicate information collected from the metadata obtained from the twitterer social profile.

### 5.5.4 System Evaluation

#### 5.5.4.1 Ground Truth

The collected and cleansed tweets of each user is carefully examined to obtain an accurate understanding of the user’s domain of interest, thereby establishing a truth dataset for developing and validating the proposed system framework at the user level. In this dataset, users are labelled and assigned to two categories: (1) **on-topic** users who show a particular interest in the political domain and (2) **off-topic** users who demonstrate no or minimal interest in the political domain. Table 2 shows a tentative list of collected users, and the actual number of users selected for the ground truth, based on an examination of all tweets.

---

8 https://blog.twitter.com/2016/announcing-an-application-process-for-verified-accounts-0.

Table 2: Ground Truth

| | #Collected users (tentative list) | Ground Truth |
|---|---|---|
| on-Topic | 356 | 268 |
| off-Topic | 378 | 314 |

The collected users of the ground truth dataset indicated in Table 2 are analysed with their historical tweets to develop the prediction model. This is used to predict the likelihood of users in the political domain.

### 5.5.4.2 Experimental Settings

The experiments for this study were carried out using RapidMiner™ software, one of the top tier design science platforms according to Gartner [76]. RapidMiner has been incorporated for conducting large scale data analytics leveraging sophisticated embedded modules that can run in-parallel inside big data environment [77, 78]. The experiment is conducted using six different prediction/classification modules, namely: Naive Bayes, Generalized Linear Model, Logistic Regression, Decision Tree, Random Forest, and Support Vector Machine.

The five machine learning techniques depicted previously were implemented, 60% of the dataset was used to train these models and the performance was computed on 40% of the dataset that was unseen for any of the implemented model optimizations. The key parameters were examined and inferred from those in the optimal models. Table 3 presents a summary of several selected hyperparameters and their settings for all of the incorporated machine learning modules.

Table 3: Selected hypeparameter settings for predictive models

| HyperParameter | Description | Value |
|---|---|---|
| Generalised Linear Model (GLM) | | |
| *Family* | Uses binomial for classification | gaussian |
| *Solver* | Used for optimisation | IRLSM |
| *Standardisation* | Standardisation numerical columns | Checked |
| *Maximum number of threads* | Controls parallelism level of building model | 1 |
| Naive Bayes (NB) | | |
| *laplace correction* | Prevents the occurrence of zero values | True |
| Logistic Regression (LR) | | |
| *Solver* | Used for optimisation | IRLSM |
| *Compute p-values* | Requests p-values computation | True |
| *Remove collinear columns* | Removes some dependent columns | True |
| *Add intercept* | Includes constant term in the model | Ture |

| Random Forest Tree (RFT) | | |
|---|---|---|
| *No. Trees* | Number of random generated trees | 100 |
| *Criterion* | On which attribute will be split | gain_ratio |
| *Max_depth* | Depth of the tree | 10 |
| Gradient Boosted Tree (GBT) | | |
| *No. Trees* | Number of generated trees | 20 |
| *maximum number of threads* | Controls parallelism level of model building. | 1 |
| *Max_depth* | Depth of the tree | 10 |
| Decision Tree (DT) | | |
| *Criterion* | On which attribute will be split | gain_ratio |
| *Max_depth* | Depth of the tree | 20 |
| *Confidence* | confidence level used for the pessimistic error calculation of pruning | 0.1 |
| *minimal gain* | The gain of a node is calculated before splitting it | 0.05 |

#### 5.5.4.2 Experimental Results

The performance of the proposed approach in this case study is inspected and reported. Five well-known different metrics, commonly used in Information Retrieval, are incorporated to measure the performance of each of the prediction module. These metrics include:

***Accuracy***: indicates the ratio of correct predictions to the total number of obtained predictions. The higher the accuracy, the better the predictive and classification model.

***Classification Error:*** refers to the ratio of incorrect predictions to the total number of obtained predictions. The lower the classification error, the better the predictive and classification model.

***Precision***: presents the ratio of correct positive predictions in comparison to all positive predictions. The higher the precision, the better the predictive and classification model

***Recall***: refers to the ratio of correct positive predictions in comparison to all positive predictions. The higher the recall, the better the predictive and classification model.

***AUC:*** indicates the area under the ROC curve. It is commonly known that the closer the model to 1 the better the model is.

Figure 3 shows the evaluation results of each metric on each prediction model. As depicted in the figure, Decision Tree algorithm overshadows other models in mostly all metrics. This is evident as Decision Tree obtains 97.95%, 2.05%, 98.57%, 96.92, and 99% in Accuracy, Classification Error, Precision, Recall, and AUC respectively. On the other hand, Support Vector Machine reports poor performance in this task in all metrics.

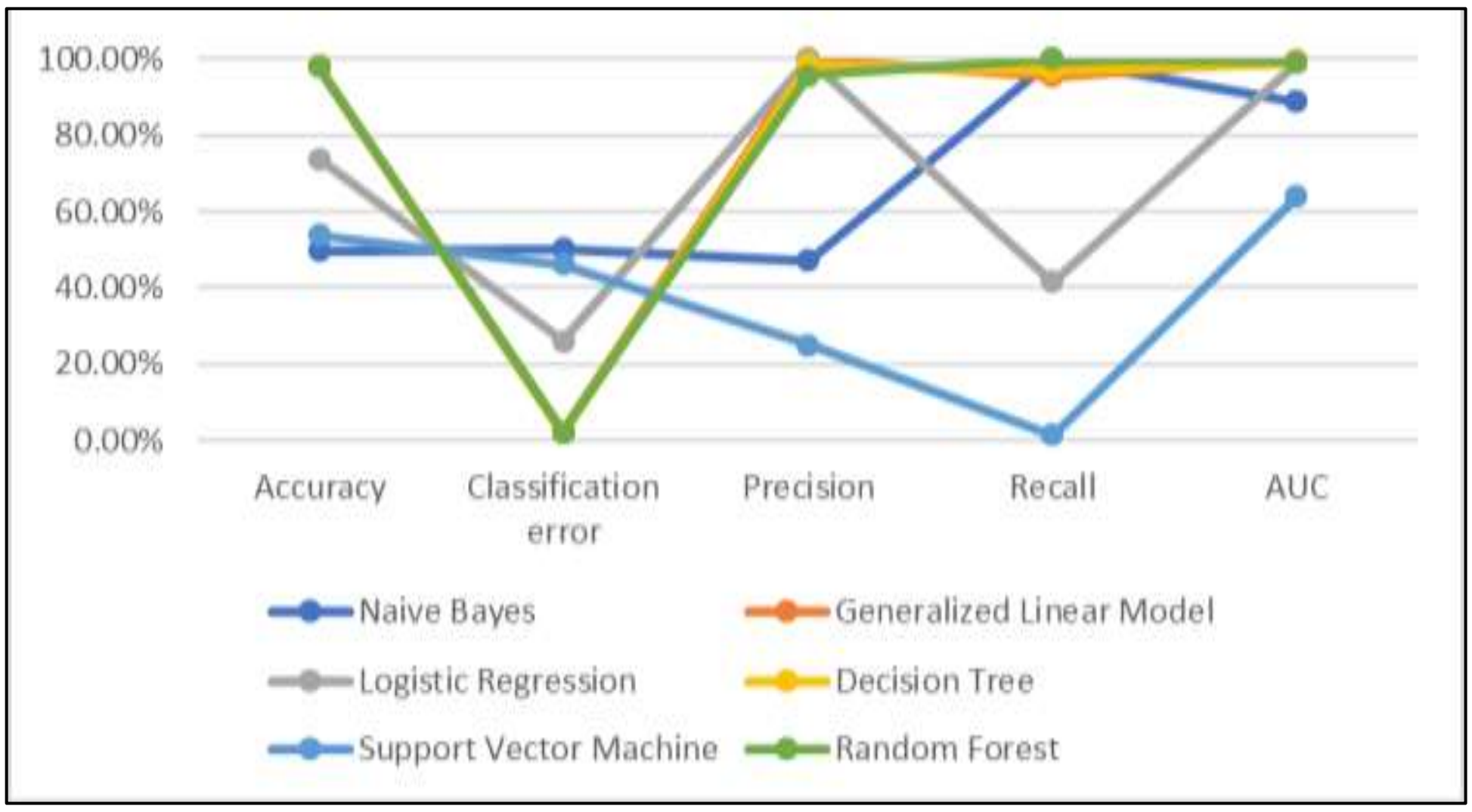

Figure 3: Evaluation performance using five different metrics on the six classifiers

Figure 3 displays the Area Under the Curve - Receiver Operating Characteristic (AUC - ROC) curves for all prediction models. The figures provide a comparison between true positives and false positives for correct and incorrect predictions. AUC-ROC curve is commonly used to scrutiny the capacity of a binary classifier model by plotting the true positive rate against the false positive rate. A good model obtains AUC that is close to 1 which indicates its ability to provide good measure of discrimination/separability. On the other hand, AUC of poor models is commonly close to 0 which indicates poor measure of separability. AUC-ROC curves shown in Figure 3 reflect the performance of each incorporated classifier in the designated task. For example, AUC-ROC curve of Support Vector Machine model depicts a poor model in this prediction task, while Generalized Linear Model, Logistic Regression and Decision Tree for example have shown high capacity of these models to distinguish between positive class and negative class.

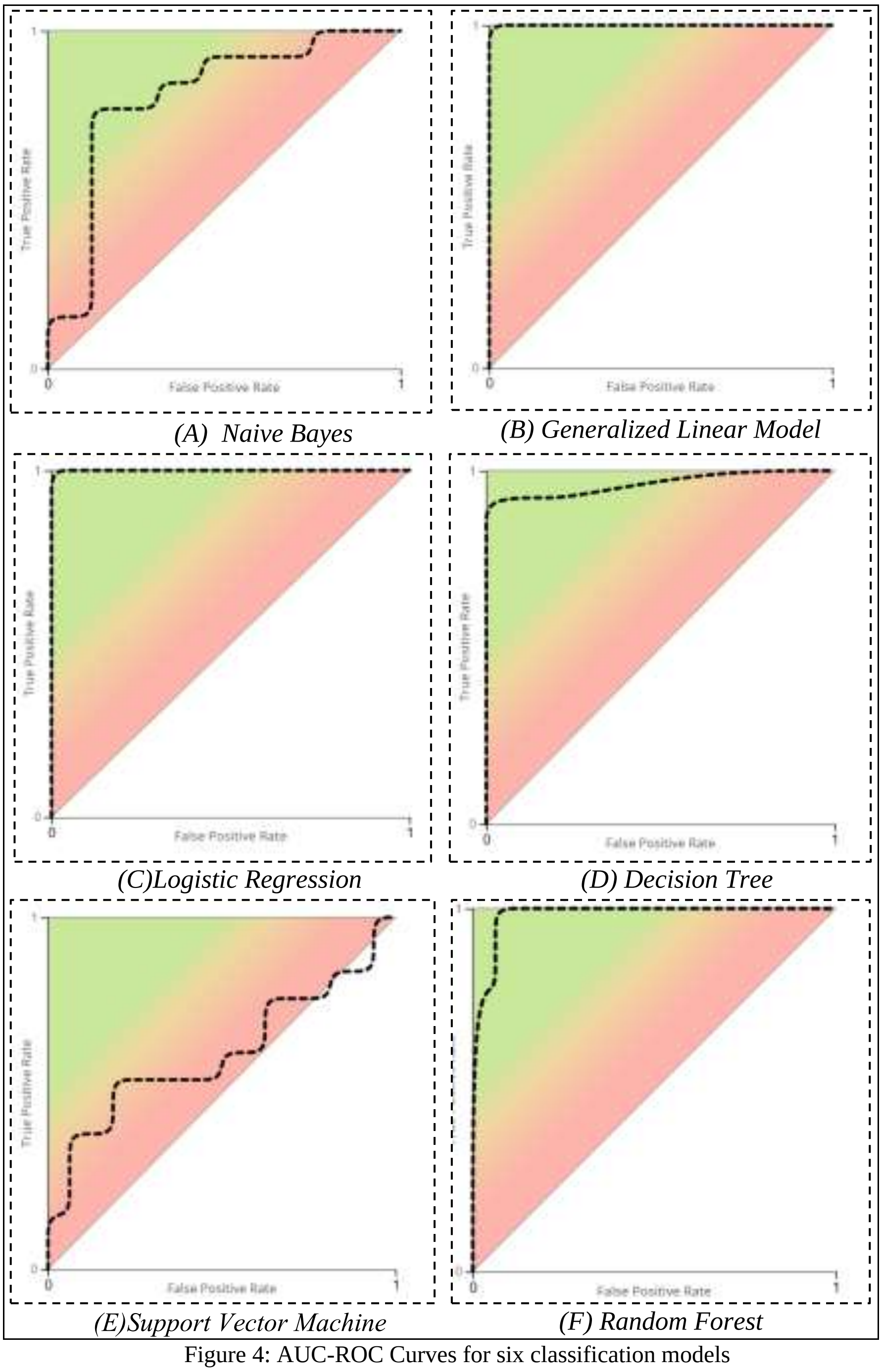

Figure 4: AUC-ROC Curves for six classification models

Figure 5 shows an example of a tree generated by the Decision Tree classifier. As depicted in the figure, *unq_politics_concept* feature is selected as a root node at which to split the tree. To evaluate this tree, we start with the root node(i.e *unq_politics_concept*) and follow the correct path through the decision nodes (in-between nodes) until we approach the leaf node. Leaf node indicates whether the user is politician (red rectangle) or non-politician (green rectangle).

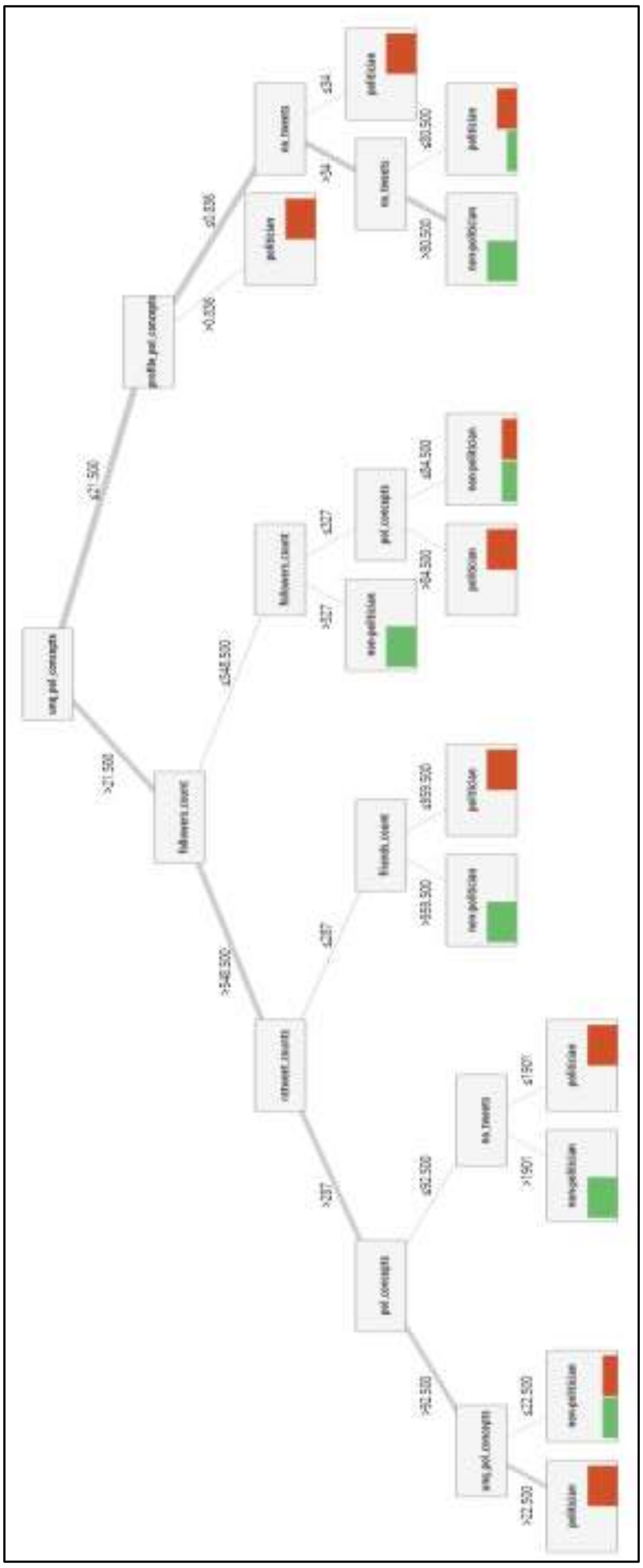

Figure 5: An Example of a generated tree from Decision Tree Prediction Models.

### 5.4.3 A Comparison with LDA and SLA

As discussed, LDA and SLA are statistically well-known models used for several topic modelling applications. In this section, an experiment is conducted to benchmark the applicability of the proposed ontology-base model at the user level against these two models, to identify a user’s main topic of interest. Gensim’s python implementation [79] of LDA and SLA is used. The collected historical tweets of two politicians’ accounts (i.e. @sarahinthesen8 and @stephenjonesALP) have been fed to the three models: LDA, SLA and the developed model incorporating a Politics Knowledge Inference. The experimental settings for LDA and SLA are set to one topic modelling, and the extracted terms indicate the 25 most contributed terms to this topic. In this approach, the top 25 frequently annotated entities from the user's tweets are extracted. Table 4 and Table 5 show the top 25 terms/entities extracted using the three approaches for @sarahinthesen8 and @stephenjonesALP respectively.

Table 4: Top entities/terms Extracted using LDA, SLA and the Developed Approach For @sarahinthesen8

| **LDA** | **LSA** | **Politics Knowledge Inference** | |
|---|---|---|---|
| | | Entity | SubType |
| refuge | refuge | Government of Australia | Organization |
| young | young | Australian Greens | Political Party |
| sarah | sarah | Member of Parliament | Politician |
| hanson | hanson | Elections | Event |
| nauru | nauru | Australian Labor Party | Political Party |
| children | children | Parliament | Organization |
| detent | detent | Liberal Party of Australia | Political Party |
| govt | govt | Malcolm Turnbull | Politician |
| australia | australia | Peter Dutton | Politician |
| green | green | Tony Abbott | Politician |
| abbott | abbott | Politics | Ontology |
| today | today | Sarah Hanson-Young | Politician |
| asylum | asylum | Electorate | Voter |
| manu | manu | Council | Organization |
| aust | aust | Politician | Person |
| people | people | inequality | Political_Slogan |
| senate | senate | Coalition | Political_Slogan |
| seeker | seeker | Joe Hockey | Politician |
| abuse | abuse | George Brandis | Politician |
| news | news | Liberal National Party of | Political Party |
| minister | time | Queensland | Political_Slogan |
| time | minister | welfare | Politician |
| dutton | dutton | Barnaby Joyce | Politician |
| turnbull | turnbull | Nick McKim | Politician |
| australian | australian | Kristina Keneally<br>Simon Birmingham | Politician |

Table 5: Top Entities/terms Extracted Using LDA, SLA and the Developed Approach For @ stephenjonesALP

| **LDA** | **LSA** | **Politics Knowledge Inference** | |
|---|---|---|---|
| | | Entity | SubType |
| illawarra | illawarra | Member of parliament | Politician |
| qt | qt | Elections | Event |
| today | today | Parliament | Organisation |
| great | great | Australian Labor Party | Political Party |
| mp | mp | Government of Australia | Organisation |
| stephen | stephen | Liberal Party of Australia | Political Party |
| good | good | Coalition | Political Slogan |
| post | post | Tony Abbott | Politician |
| school | school | Council | Organisation |
| abbott | abbott | Anthony Albanese | Politician |
| jone | jone | Politics | Ontology |
| photo | photo | Julia Gillard | Politician |
| auspol | auspol | Electorate | Voter |
| parliament | day | Greg Combet | Politician |
| day | jame | Sharon Bird | Politician |
| jame | parliament | Joe Hockey | Politician |
| big | big | Mark Butler | Politician |
| support | support | Malcolm Turnbull | Politician |
| nbn | nbn | Kate Ellis | Politician |
| house | house | Barack Obama | Politician |
| facebook | facebook | Joel Fitzgibbon | Politician |
| time | time | Jamie Briggs | Politician |
| fb | fb | Australian Greens | Political Party |
| australia | australia | Steven Ciobo | Politician |
| purser | purser | Greg Hunt | Politician |

The list of the top contributed terms identified using 1-topic modelling for each user incorporating LDA and SLA illustrates the inadequacy of these approaches in identifying a high-level topic. On the other hand, with the top 25 entities annotated for both users using the developed approach, the high-level topic (i.e. politics) is highly noticeable. In the developed proposed system framework, each entity is linked with a specific class in the ontology. The knowledge obtained for each entity can be enriched to facilitate the overall semantic interlinking which leads to a better understanding of the domain of knowledge. Interlinking and enrichment are not applicable to LDA and SLA. Furthermore, all the top entities annotated using the developed proposed system framework indicate politics entities, although some of the most frequently occurring terms extracted using LDA and SLA are political entities. In a nutshell, the outcome of this experiment shows the applicability and effectiveness of the developed proposed approach.

## 5.6 Conclusion and Future Works

This chapter provides a brief on predictive analytics and its usage in the context of SBD. Further, the commonly used machine learning algorithms to develop predictive models in order to predict new social situations are presented and discussed. The predictive model represents the relationship between predictive features to social situations. We discuss the following machine algorithms namely logistic regression (LR), generalized linear model, naïve bayes (NB), tree based classification, random forest, gradient boosted tree and deep learning. An array of several applications, tools and APIs that took advantage of these advanced machine learning algorithms are presented. Finally, an experiment that incorporate semantic web technologies and predictive analytics are implemented and its utility is demonstrated.

The following are the possible enhancements and research directions to be addressed in the future work:

(i) Besides politics, a domain-based analysis of several domains of knowledge will be conducted to acquire a more comprehensive insight into each domain. This is to facilitate the development of several domain-based ontologies leveraged by semantic web technologies and Linked Open Data.

(ii) Machine learning will be utilised to achieve the abovementioned research objectives through multi-classification applications, to predict the likelihood of user interest in several domains of knowledge.

(iii) Mobile cloud computing technologies have good potential for the future of social media. Mobile devices such as iPhones, tablets, laptops, smartphones etc can be connected to the Internet. The Internet of Things can capture all social media data. The centre cloud using the machine learning algorithm, such as deep learning, can be harnessed to analyse people's needs and behaviours inferred from social data stored in the cloud.

(iv) The current machine learning approaches assume that uncertainty and incompleteness do not significantly affect the accuracy of the Twitter classification. In fact, data uncertainty and incompleteness may exist. In the future, we will formulate the data uncertainty and incompleteness as fuzzy numbers which can be used to address imprecise, uncertain and vague data. Based on the fuzzy numbers, a fuzzy-based machine learning algorithm will be developed to estimate the effect of data uncertainty and incompleteness.